\- **Josephson Tunneling and Nanosystems**

Yurii N. Ovchinnikov[a,b] and Vladimir Z. Kresin[c]

[a)] L.D.Landau Institute for Theoretical Physics, Russian Academy of Sciences, 117334, Moscow, Russia

[b)] Max-Planck Institute for Physics of Complex Systems Dresden, D-01187, Germany

[c)] Lawrence Berkeley Laboratory, University of California at Berkeley, CA 94720

**Abstract**

Josephson tunneling between nanoclusters is analyzed. The discrete nature of the electronic energy spectra, including their shell ordering, is explicitly taken into account. The treatment considers the two distinct cases of resonant and non-resonant tunneling. It is demonstrated that the current density greatly exceeds the value discussed in the conventional theory. Nanoparticles are shown to be promising building blocks for nanomaterials-based tunneling networks.

## I. **Introduction**

The electronic states in many metal clusters form energy shells similar to those in atoms and nuclei (see ,e.g., the reviews [1-3]). This shell structure implies the presence of orbital degeneracy 2(2l+1), where l is the orbital angular momentum. It was shown in our paper [4] that this fundamental feature leads, under special but realistic conditions, to a great strengthening of pair correlations and, consequently, to a drastic increase in the value of the critical temperature of the superconducting transition.

In the paper [4] we focused on pair correlations in an isolated nanocluster. This phenomenon is analogous to pairing in atomic nuclei (see, e.g., the reviews [5],[6]). For such a case, pair correlation manifests itself in the cluster spectrum: pairing leads to an increase of the spacing (at $T<T_c$) between the highest occupied level and lowest unoccupied energy levels, odd-even effects, etc. A measured jump in the heat capacities of $Al_{45}^-$ and $Al_{47}^-$ clusters at $T_c \approx 200K$ [7] was the first observation of the phenomenon; the amplitude and width of the jump ,as well as its position, are in good agreement with the theory [4].

The present paper is concerned with charge transfer (tunneling) between superconducting nanoclusters and can be viewed as a continuation of the study [ 4 ]. This problem is of definite interest because it is directly related to possibility of building tunneling networks. Such networks can be used to transport the *macroscopic* superconducting current arising from Josephson coupling between the network-forming clusters.

The problem of Josephson coupling between two nanoparticles has been studied in the interesting paper [8]. Its authors stressed an important point: although the phase of the order parameter for an isolated nanoparticle is not defined (since it has a fixed number of electrons), it is nevertheless possible to define a phase *difference*, which is the quantity that enters the expression for the Josephson current. The paper [8] contains a calculation of the Josephson energy. However, its authors consider the case of a continuous electronic energy spectrum and their treatment contains an integration over the energy. Such an approximation is sufficient for large nanoparticles, but for the nanoclusters with $N \approx 10^2$-$10^3$ (N is the number of delocalized electrons), the discrete nature of the electronic spectrum is a key factor that strongly impacts the charge transfer phenomenon. This factor will be explicitly taken into account in the following.

## II. Current

Consider two superconducting clusters separated by a tunneling barrier (Fig. 1 ).We assume that the clusters are embedded in a matrix or placed on a substrate. As was noted above, the specifics of the system studied here is that, contrary to usual bulk superconductors, the clusters are characterized by discrete energy spectra . Because of this, Josephson tunneling between them needs to be analyzed with considerable care. For example, in contrast to the bulk case, one should take into account the fact that tunneling itself splits the energy levels and leads to the formation of symmetric and antisymmetric terms ( see, e.g., [9]).

The phenomenon is usually described by the tunneling Hamiltonian formalism (see, e.g., [10] and the reviews [11,12]). However, here we will start from a more general expression. Indeed, first of all it is not obvious that the tunneling Hamiltonian method is applicable to the case of strongly quantized systems. In addition, and this is even more important, there exists a very interesting case ("resonant" channel, see below) when the result turns out to be different from that obtained by using the tunneling Hamiltonian.

Let us start with the general expression for the current:

$$\vec{j} = \{(ie\hbar/m)[\left(\partial/\partial\vec{r}^{'}\right) - \left(\partial/\partial\vec{r}\ \right)] - (2e^2\hbar\vec{A}/m)\}G(\vec{r},\tau;\vec{r}^{'},\tau^{'})_{|\vec{r}=\vec{r}^{'};\tau^{'}=\tau+0} \qquad (1)$$

where $G(\vec{r},\tau;\vec{r}^{'},\tau^{'})$ is the electronic Green's function . The Green's function G describes a whole system containing both nanoparticles (left L and right R, see fig.1). The electronic wave functions of the nanoparticles may overlap, which reflects the possibility of tunneling between them. The $\hat{G}$-matrix

$$\begin{pmatrix} -G & F \\ F^{+} & \tilde{G} \end{pmatrix}$$

satisfies the equation:

$$\begin{pmatrix} (\partial/\partial\tau)+\hat{H} & \Delta \\ \Delta^{*} & (\partial/\partial\tau)-\hat{H} \end{pmatrix}\hat{G} = \delta(\vec{r}-\vec{r}^{'})\delta\left(\tau-\tau^{'}\right) \qquad (2)$$

Here F is the pairing Green's function [13], $\Delta \equiv \Delta(\vec{r})$ is the order parameter, and the Hamiltonian is

$$\hat{H} = -\frac{\hbar^2}{2m}\partial^2/\partial\vec{r}^{\,2} + U(r) - \mu \qquad (3\ \ )$$

U is the total potential energy and $\mu$ is the chemical potential whose value depends on the temperature and the number of electrons. We employ the thermodynamic Green's functions formalism ( see, e.g., [14 ]), so that $\tau$ is the imaginary "time".

Let us introduce also the matrix operator $\hat{K}$ and its eigenfunctions defined as

$$\hat{K} = \begin{pmatrix} \hat{H} & \Delta \\ \Delta^* & -\hat{H} \end{pmatrix}; \hat{f} = \begin{pmatrix} f_1 \\ f_2 \end{pmatrix} \tag{4}$$

so that

$$\hat{K}\hat{f} = E\hat{f} \tag{5}$$

The explicit expression for the $\hat{f}$-functions depends on the potential energy U (see Eq.(3)). Based on Eqs.(2),(3) and (5),one can express the G-matrix in terms of $\hat{f}$-functions , obtaining:

$$\hat{G} = T\sum_{\omega_n}\sum_{E}\frac{e^{-i\omega_n(\tau-\tau')}}{-i\omega_n + E}\hat{f}(\vec{r})\hat{f}^+(\vec{r}') \tag{6}$$

Expression ( 6 ) contains summation over all energy branches.

Let us introduce the functions $\hat{f}^{L}$ and $\hat{f}^{R}$ which describe isolated L and R clusters. The functions $\hat{f}^{L}$ and $\hat{f}^{R}$ satisfy Eq.(5) with U corresponding to L and R clusters,respectively. Considering separately the right (R) and left (L) nanoparticles, one obtains:

$$\hat{f}^{i} = \begin{pmatrix} 1 \\ \dfrac{\Delta^{i^*}}{E^{i}+\xi^{i}} \end{pmatrix} f^{i} \Big/ g^{i} \qquad (7)$$

$$i \equiv \{L;R\}$$

Here

$$E^{i} \equiv E^{i^{\pm}} = \pm\left[(\xi^{i})^{2}+(\Delta^{i})^{2}\right]^{1/2}; \xi^{i} = E_{n}^{i}-\mu \qquad (8)$$

$E_{n}^{i}$ and $f^{i}$ correspond to the normal state of the nanoparticles, $E^{i}$ is the excitation energy for the superconducting state, and $g^{i}$ are the normalization coefficients:

$$g^{i2} = \left(1+\frac{|\Delta^{i}|^{2}}{\left(E^{i}+\xi^{i}\right)^{2}}\right)\int d\vec{r}\,|f^{i}|^{2} \qquad (9)$$

It is important that tunneling results in the states of initially isolated nanoparticles becoming mixed and the energy levels $E^L$ and $E^R$ becoming split. The splitting is determined by the matrix element of the tunneling parameter, $\gamma_{\upsilon\upsilon_1}$ [$\nu$ and $\nu_1$ are quantum numbers for the L and R clusters, so that $\nu\equiv\nu(L)$ and $\nu_1\equiv\nu_1(R)$]. The expression for $\gamma_{\upsilon\upsilon_1}$ can be obtained from the Eq.(5). A calculation (see Appendix A) leads to the expression

$$\gamma_{\nu\nu_1} = (\hbar^{2}/2m)\int_{S} d^{2}S\vec{n}(g_{\upsilon}^{L}g_{\upsilon_1}^{R})^{-1}\times[(f_{\nu_1}^{R})^{*}\left(\partial f_{\nu}^{L}/\partial\vec{r}\right)-f_{\nu}^{L}(\partial f_{\nu_1}^{R}/\partial\vec{r})^{*}]$$
$$\times[1-\Delta^{R}\Delta^{L^*}\left(E_{\upsilon}^{L}+\xi_{\upsilon}^{L}\right)^{-1}\left(E_{\nu_1}^{R}+\xi_{\upsilon_1}^{R}\right)^{-1}]$$

(10)

The quantities $g_v^i$ and $E_v^i$ are defined by Eqs.(8),(9).
The expression for the splitting can be obtained from the secular equation and has a form:

$$E_{1,2} = 0.5\{(E_v^L + E_{v_1}^R) \pm (\Delta E)\left[1 + 4 \,|\, \tilde{\gamma}_{vv_1} |^2\right]^{1/2}\}$$

(11)

Here

E$^L$≡ $E_\upsilon^L$, E$^R$≡ $E_{\upsilon_1}^R$ , ΔE=$E_\upsilon^L$-$E_{\upsilon_1}^R$, $\tilde{\gamma}_{\upsilon\upsilon_1} = \gamma_{\upsilon\upsilon_1}(E_\upsilon^L - E_{\upsilon_1}^R)^{-1}$. (11')

The most interesting case corresponds to close values of $E_\upsilon^L$and $E_{\upsilon_1}^R$. We assume also that the energy level for an isolated cluster, e.g., $E_\upsilon^L$, is not degenerate and its distance from the neighboring level ,δE, is such that δE>>|$\gamma_{\upsilon\upsilon_1}$|. A more general case is straightforward to consider and present a similar general picture, albeit with a more complicated secular equation.

With use of Eqs.(7)-(11), one can write out detailed expression for the functions $\Psi_1$ and $\Psi_2$ which form the general normalized solution of Eq.(5) , namely,

$$\hat{\Psi}_{1;v} = \hat{f}_\upsilon^L + \tilde{\gamma}_{vv_{1m}}^{1,R} \hat{f}_{\upsilon_{1m}}^R + \sum_{\pm v_1, v_1 \neq v_{1m}} \tilde{\gamma}_{vv_1}^{1,R} \hat{f}_{\upsilon_1}^R - \sum_{\pm \tilde{v}, \tilde{v} \neq v_{1m}} \tilde{\gamma}_{vv_{1m}}^{1,R} \tilde{\gamma}^*_{\tilde{v}v_{1m}} \hat{f}_{\tilde{\upsilon}}^L$$

$$\hat{\Psi}_{2;v_1} = \hat{f}_{\upsilon_1}^R + \tilde{\gamma}_{v_m v_1}^{2,L} \hat{f}_{\upsilon_m}^L - \sum_{\pm \upsilon, v \neq v_m} \tilde{\gamma}^*_{vv_{1m}} \hat{f}_\upsilon^L + \sum_{\pm \tilde{\upsilon}_1, \tilde{\upsilon}_1 \neq v_m} \tilde{\gamma}_{v_m v_1}^{2,L} \tilde{\gamma}_{v_m \tilde{v}_1} \hat{f}_{\tilde{\upsilon}_1}^R \qquad (12)$$

*Here*

$$\tilde{\gamma}_{vv_1}^{1,R} = \gamma_{vv_1}(E_{1;v\upsilon_{1m}} - E_{\upsilon_{1m}}^R)^{-1}; \tilde{\gamma}_{vv_1}^{2,L} = \gamma_{vv_1}^*(E_{2;\upsilon\upsilon_1} - E_\upsilon^L)^{-1}$$

(12')

$\gamma_{\nu\nu_1}$ and $E_1, E_2$ are defined by Eqs.(10) and (11),respectively.

The notation $\nu_{1;m}$ denote the state whose energy value is closest to the value described by the quantum number $\nu$. The essential fact is that because of tunneling we always deal with the pair of states $\nu, \nu_1$ corresponding to the L and R nanoclusters. Based on Eqs.(7) and (12), we can write down the following expression for the eigenfunctions $\hat{f}$ (see Eqs. ( 5) and (6)):

$$\hat{f} = \{\hat{\Psi}_{1;\nu}[1 + |\tilde{\gamma}^{1,R}_{\nu\nu_{1m}}|^2]^{-1/2}; \hat{\Psi}_{2;\nu_1}[1 + |\tilde{\gamma}^{2,L}_{\nu_m\nu_1}|^2]^{-1/2}\} \quad (13)$$

Based on Eqs.(12 ),(13) and Eqs.(7)-( 9 ) one can write out the expression for the Green's function (6),and then use this expression to evaluate the current (1).

After long but straightforward calculation we arrive at the following expression for the Josephson current:

$$j = j_m \sin\alpha \quad (14)$$

where $\alpha$ is the phase difference, and

$$j_m = j_m^{(1)} + j_m^{(2)} \tag{15}$$

$$j_m^{(1)} = \frac{e\hbar^3}{m^2} T \sum_{\omega_n} \sum_{\nu,\nu_1} | T_{\nu,\nu_1} |^2 \frac{|\Delta^L||\Delta^R|}{\left[\omega_n^2 + (E_\nu^L)^2\right]\left[\omega_n^2 + (E_{\nu_1}^R)^2\right]}$$

$$j_m^{(2)} = \frac{e\hbar^3}{m^2} \sum_{\omega_n} [ \sum_{\nu \neq \tilde{\nu}} | T_{\tilde{\nu},\nu_{1m}} |^2 \frac{|\Delta^L||\Delta^R||\tilde{\gamma}^{1,R}_{\upsilon\upsilon_{1m}}|^2}{\left(\omega_n^2 + E^2_{1;\nu\nu_{1m}}\right)\left((E^L_{\tilde{\nu}})^2 - (E^R_{\nu_{1m}})^2\right)\left(1 + |\tilde{\gamma}^{1,R}_{\upsilon\upsilon_{1m}}|^2\right)} -$$

$$\sum_{\nu_1 \neq \tilde{\nu}_1} | T_{\nu_m,\tilde{\nu}_1} |^2 \frac{|\Delta^L||\Delta^R||\tilde{\gamma}^{2,L}_{\upsilon_m\upsilon_1}|^2}{\left(\omega_n^2 + E^2_{2;\upsilon_m\nu_1}\right)\left((E^L_{\upsilon_m})^2 - E^R_{\tilde{\nu}_1}\right)\left(1 + |\tilde{\gamma}^{2,L}_{\upsilon_m\upsilon_1}|^2\right)} - \sum_{\nu,\nu_1;\nu_1 \neq \nu_{1m}} | T_{\nu\nu_1} |^2 \frac{|\Delta^L||\Delta^R|}{\left[\omega_n^2 + (E_\nu^L)^2\right]\left[\omega_n^2 + (E_{\nu_1}^R)^2\right]} \frac{|\tilde{\gamma}^{1,R}_{\upsilon\,\upsilon_{1m}}|^2}{1 + |\tilde{\gamma}^{1,R}_{\upsilon\,\upsilon_{1m}}|^2} \; ]$$

(16)

Here

$$| T_{\nu,\nu_1} |^2 = | \int_S d\vec{S} \left[ f_{\nu_1}^{R*} \left( \partial f_\nu^L / \partial \vec{r} \right) - f_\nu^L (\partial f_{\nu_1}^R * / \partial \vec{r}) \right] |^2 \left( \int d\vec{r} \, | f_\nu^L |^2 \right)^{-1} \left( \int d\vec{r} \, | f_{\nu_1}^R |^2 \right)^{-1} \tag{17}$$

The quantities $\tilde{\gamma}^{1,R}_{\upsilon\,\upsilon_1}$ and $\tilde{\gamma}^{2,L}_{\upsilon\,\upsilon_1}$ are defined by Eq.(12') . The expression for the amplitude $j_m$ contains also the term $j_m^{(3)}$ which is of higher order in the tunneling parameter $\gamma_{v\,v_1}$. This term is usually smaller than $j_m^{(2)}$ and we shall not write it down here.

Note also that the term $T_{v,v_1}$ ,Eq.(17), is the matrix element of the tunneling Hamiltonian. One can see directly from Eqs.(10),(17) that $T_{v,v_1}$ is related to the tunneling parameter introduced in Eq.(10). Namely, $|T_{v,v_1}|^2 = (2m/\hbar^2)\gamma^2_{v;v_1}$ ($\Delta^i$=0). It is interesting that that the usual tunneling matrix element appears in Eqs.(16),(17) as a consequence of the calculation based on the general equation (1), but not as the starting point. For the conventional case of continuous spectra we obtain the usual expression (see below,Eq.(18)). Therefore, Eqs.(16),(17) represent a generalization of the usual expression, obtained by the tunneling Hamiltonian formalism, to the case of Josephson transfer between systems with discrete energy spectra.

The expression for $j_m^{(1)}$ contains the product $F^L F^R$ which appears in the tunneling Hamiltonian formalism ( see,e.g., [ 11 ]); $F^i$ is the pairing thermodynamic Green's function (see,e.g.,[14]), e.g., $F^L = \Delta^L[\omega_n^2 + (E_\nu^L)^2]^{-1}$ ; the tunneling matrix element has a form (17 ). Nevertheless, the result is different, because the summation can not be replaced by integration.

Note that in the absence of resonant levels : $(E_\nu^L - E_{\upsilon_1}^R) >> \gamma_{\nu\nu_1}$ one can neglect the term $j_m^{(2)}$ , since $\tilde{\gamma}_{\nu\nu_1} << 1$. However, for the case of resonance (when $\tilde{\gamma}_{\upsilon\,\upsilon_1}^{1,R} = \tilde{\gamma}_{\upsilon\,\upsilon_1}^{2,L}$ =1) the situation is different (see below,Sec.III). Then the expression for the amplitude of the Josephson current becomes entirely different from that obtained with the tunneling Hamiltonian formalism.

If the summation over the quantum numbers can be replaced by integration (cf.[ 8 ]) ,we obtain with use of Eq.(16) the well-known expression for the amplitude of the Josephson current [10 ]:

$$j_m^{conv} = \frac{\pi\Delta}{2eR_{conv}} th\left(\Delta/2T\right) \quad where \quad R_{conv}^{-1} = \nu_F^2 \,|T_{conv}|^2 \qquad (18)$$

Here $\nu_F$ is the density of states, $\nu_F = mp_F V/\pi^2 = 3(\Delta\bar{E})^{-1}$ ; $\Delta\bar{E} = E_F/N$ is the average energy spacing, N=nV is the number of free electrons,V is the volume of the nanoparticle, and n is the carrier concentration.

However, if the electronic states form an energy shell pattern, the picture is very different. The states are classified by their orbital angular momentum l, and possess the degeneracy ( g=2(2l+1)). Because of this feature,it is no longer permissible to replace summation by integration.

Let us consider the non-resonant and resonant cases.

**III. Non-resonant and resonance cases.**

Non-resonant case. For the non-resonant situation which is rather typical ,the amplitude of the Josephson current is determined by the term $j_m^{(1)}$ which is described by Eq.(16). Note that the term $j_m^{(2)}$ can be simplified even for some resonant cases. For example, if we are dealing with two similar "magic" clusters ,there will be a cancellation of the first two terms entering $j_m^{(2)}$ .

As was noted above, the expression for $j_m^{(1)}$ ,Eq.(16) is of the form that follows from the tunneling Hamiltonian formalism. Nevertheless, because of discrete nature of the spectrum, the result turns out to be different from that for usual superconductors.

As an example, let us consider two identical “magic” clusters . The main contribution comes from the highest occupied (HOS) and the lowest unoccupied (LUS) shells (cf.[4 ]). In the low temperature region (T→0 K) ,the summation over $\omega_n$ can be replaced by integration $\left(2\pi T\sum_{\omega_n}\rightarrow\int d\omega\right)$. In addition, one should take into account the dependence $\Delta(\omega)$.

With use of Eq.( 16) , and performing the integration over $\omega$ , we obtain the following expression for the amplitude of the current (at T→0K):

$$j = \frac{e\hbar^3}{4m^2}\sum_{\nu\nu'=H,L} |\hat{T}_{\nu\nu'}|^2 |\Delta|^2 W^{-3}$$

$$\text{where} \qquad (19)$$

$$W \approx [|\Delta|^2 + (\Delta E_{LH}/2)^2]^{1/2}$$

Here $\hat{T}_{\nu\nu'}$ is the tunneling matrix element (see Eq.(17 ) and Appendix ), and $\Delta E_{LH}$ is the energy space between the HOS and LUS shells . For simplicity, let us neglect the dependence $\Delta(\omega)$.Assume a distance of d≈15A.

Based on Eq.( 16),we obtain:

$$j_m = 6 \bullet 10^{-6} \frac{e\hbar^3 |\Delta|^2}{m^2 a^4} W^{-3} \qquad (20)$$

If we calculate $j_m$ with use of usual semiclassical picture (Eq.(18 )),we arrive at the following expression for the ratio of the current amplitudes for the cases of nanocluster vs. a usual superconductor $\eta = (j_m^{cl} / j_m^{sc})$

$$\eta \approx 6 \bullet 10^2 (\hbar^2 / ma^2)^3 (\Delta_\infty E_F)^{-1} \Delta^2 W^{-3} \qquad (21)$$

Let us consider the specific case studied in [ 4 ], namely, a cluster with the following realistic set of parameters:

$\Delta E$ = 65 meV, $\tilde{\Omega}$=25 meV, $m^*$=0.75$m_e$,

$k_F$ = 1.5x$10^8$cm$^{-1}$, the radius R = 6A,

and $G_H$= 30 ;$G_L$ = 18 (e.g., $l_H$=7, $l_L$=4) (22)

With these values , we obtain

$$\eta \approx 10^2 \qquad (23)$$

Therefore, the discreteness of the electronic spectrum and the presence of shell structure result in the amplitude of the Josephson current between two nanoclusters greatly exceeding that obtained for the usual junction.

Consider a different case, namely a Josephson contact with two $Al_{46}$ clusters, so that N=138. This case is interesting because clusters with N of a similar magnitude have been observed to display a jump in the heat capacity at $T_c \approx 200$ K [7]. The cluster $Al_{46}$ is characterized by the following set of parameters: $a \approx 5.5$A, $l_{HOS}=1$, $l_{LUS}=7$, $E_0 \approx 1.5.10^{-11}$erg, $\delta U_0 \approx 5$eV; assume, as above, that the distance d=15A. According to the data [ 15 ], $\Delta E \approx$ 40meV. One can show that the key contribution comes from the HOS and LUS states. With use of these parameters, we find that $\eta \approx 5.10^2$.

We believe that $Al_{46}$ clusters (or $Al_{45}^-$ ions) with d < 15A may represent a good choice for experimental work, because the preparation and spectroscopy of Al clusters is a well developed technique (see, e.g.,[1,7,15 ]).

<u>Resonant</u> <u>case.</u> In this case $E^L_\upsilon = E^R_{\upsilon_1}$ , and correspondingly $\tilde{\gamma}^{1,R}_{\upsilon\,\upsilon_1} = \tilde{\gamma}^{2,L}_{\upsilon\,\upsilon_1} = 1$ (see Eqs.(11) and (12')).Then the expression for the current can be written as the following sum:

$j_{m;res} = j_1 + j_2$

$$j_1 = \frac{e\hbar^3}{2m^2} T \sum_{\omega_n} \sum_{\nu,\nu_1} |T_{\nu,\nu_1}|^2 \frac{|\Delta^L||\Delta^R|}{\left[\omega_n^2 + (E^L_\nu)^2\right]\left[\omega_n^2 + (E^R_{\nu_1})^2\right]}$$

$$j_2 = \frac{e\hbar^3}{2m^2} T \sum_{\omega_n} \{ \sum_{\nu \neq \tilde{\nu}} |T_{\tilde{\nu},\nu_{1m}}|^2 \frac{|\Delta^L||\Delta^R|}{\left(\omega_n^2 + E^2_{1;\nu\nu_{1m}}\right)\left((E^L_{\tilde{\nu}})^2 - (E^R_{\nu_{1m}})^2\right)} - \sum_{\nu_1 \neq \tilde{\nu}_1} |T_{\nu_m^{\cdot},\tilde{\nu}_1}|^2 \frac{|\Delta^L||\Delta^R|}{\left(\omega_n^2 + E^2_{2;\upsilon_m\nu_1}\right)\left((E^L_{\upsilon_m})^2 - E^R_{\tilde{\nu}_1}\right)} \} \quad (24)$$

The expression for $j_2$ in Eq. ( 24 ) contains two terms, and the situation should be approached with considerable care. For example, as was noted above, these two terms cancel if the junction is composed of two identical spherical clusters.For clusters with incomplete shells (e.g. those with a prolate configuration) one needs to properly select the mutual orientation of the identical clusters. Then the $j_2$ term becomes dominant, and one should expect an even greater increase in the amplitude of the current. Indeed, Eqs.(24) show that the ratio $j_2/j_1$ contains the parameter $\Delta / \delta\varepsilon >> 1$, where $\Delta$ is the energy gap and $\delta\varepsilon$ is the energy spacing in the absence of pairing.

For example, for the resonant channel to be dominant, one can select prolate clusters with perpendicular orientation. Then the amplitude of the current is determined by the matrix element T{l.m.n;l,0,n), see Appendix B. Consider, for example, a junction formed by two $Al_{56}^-$ clusters. Such a cluster has a slightly deformed prolate configuration. We can use the expression for $j_m^{(2)}$ (see(16 )) As a result, after long, but straightforward calculation, we obtain: $\eta \approx 10^3$. This value is an order of magnitude larger than Eq.(23).

Therefore, indeed, the presence of the resonant channel leads to even greater increase of the current amplitude.

## IV. **Discussion. Tunneling networks**.

The discussion of Josephson tunneling implies that the clusters are positioned on some substrate. This requires "soft landing," i.e., it is essential that the cluster-substrate interaction not extinguish the electronic shell structure. As is known, this is a serious experimental challenge, but there has been notable progress in this field with use of $C_{60}$ – based substrate [16] ( see also [17]) , so one can anticipate that future work will solve this problem.

The amplitude of the current strongly depends on the distance between the electrodes (in our case, the shortest distance between the clusters, Fig.1). An increase in the distance *d* (e.g., to $d\approx 20$Å, see, e.g., [18]) will decrease the Josephson energy, and the Coulomb blockade effect can became essential. But for the relatively short distance $d\approx 5$ Å the value of the Josephson energy $E_J=hI_c/2e$ greatly exceeds that of $E_c$. Indeed, for junctions formed by Al clusters with the parameters $d\approx 5$ Å, $\rho\approx 7$ Å, $\varepsilon\approx 10$, barrier height $U\approx 1$eV,we obtain $E_J\approx 3$eV (near T=0K); this value greatly exceeds $E_c\approx 0.2$ eV.

In this paper we have focused on the properties of a single Josephson junction formed by two nanoclusters. Such junctions can be used to build cluster-based tunneling networks. In principle, the design of such networks can proceed in two directions. First of all, a network can be formed by placing the nanoclusters on a surface. The tunneling chain can form via the percolation scenario. A similar experiment has been performed for larger nanoparticles in [19]. If the number of deposited clusters increases, then eventually the percolation threshold can be crossed, and there appears macroscopic charge transfer through the Josephson tunneling chain. Such a scenario requires that the cluster shell structure be preserved upon their surface deposition. As noted above, recent progress with “soft landing” techniques looks quite promising in this regard.

The present analysis shows that the construction of superconducting networks out of superconducting nanoparticles is a promising direction. The properties of such a network will be discussed in more detail elsewhere.

Another version is a cluster-based 3D crystal, with superconducting current caused by Josephson tunneling between neighboring clusters . Such a crystal was introduced theoretically in [20]. An example of such a crystal, formed by $Ga_{84}$ clusters and superconducting with $T_c \approx 8K$, has been described in Refs. [21,22 ]. One can foresee customized crystal synthesis out of different clusters (e.g., $Ga_{56}$, see [4]), leading to higher values of Tc.

In summary, we have investigated the Josephson contact between two nanoclusters. This work is a continuation of the study [4] which described the pairing state in individual nanoclusters. The discrete nature of the electronic spectrum and the presence of energy shell ordering make a strong impact on the amplitude of the Josephson current. For the system of interest, this amplitude may greatly exceed the values in usual superconductors. In addition, the peculiar case of resonant tunneling can be observed.

The research of VZK was supported by AFOSR. The research of YNO is supported by EOARD.

## Appendix

A. In order to calculate the matrix element of the tunneling operator ,we start with equation

$$(\hat{K}-E)\hat{\Psi}_1 = 0 \qquad \hat{\Psi}_1 = \begin{pmatrix}\psi_1\\ \psi_2\end{pmatrix} \quad . \qquad \text{(A.1)}$$

(cf.Eq.(5)).We obtain:

$$\int_{R.} d\vec{r}\left(1;\Delta^R\ (E^R_{\nu_1}+\xi^R_{\nu_1})^{-1}\right)\left(f^{R^*}_{\nu_1}/g^R_{\nu_1}\right)(\hat{K}-E)\hat{\Psi}_1 = 0 \qquad \text{(A.2)}$$

With use of expression (5), A.2) and the relation:

$$f^*\left(\partial^2/\partial r^2\right)\psi\ = \partial/\partial r(f^*\partial\psi\ /\partial r - \psi\ \partial f^*/\partial r + \psi\ \partial^2 f^*/\partial r^2$$

after a straightforward calculation, we arrive at expression in Eq. (10)

B. Let us describe also the evaluation of the tunneling matrix element, see Eq.(17). For a spherical cluster the function $f_\nu$ can be written in the form:

$$f_\nu = \mathrm{Y}_l^m J_{l+1/2}(kr)/(kr)^{1/2}(r<a)$$
$$and$$
$$f_\nu = C\mathrm{Y}_l^m K_{l+1/2}(pr)/(pr)^{1/2}(r>a) \qquad \text{(B.1)}$$

Here $\mathrm{Y}_l^m$ , $J_{l+1/2}$, and $K_{l+1/2}$ are spherical and Bessel functions, $k=(2mE_0)^{1/2}$,$p=[2m\delta U_0]^{1/2}$ ,$\delta U_0=(\delta U - E_0)$, $\delta U$ is the height of the barrier, a is the cluster radius. Expressions (B.1) can be used also as a first approximation for slightly deformed clusters. The constant C can be determined with use of the usual boundary conditions at r=a and is equal to:

$$C = -(E_0 / \delta U_0)^{1/4} J_{l-1/2}(ka)[K_{l-1/2}(pa) + (l+1)K_{l+1/2}(pa)/(pa)]^{-1}$$

For the almost identical clusters ,we obtain:

$$|T_{\nu\nu_1}|^2 = (d/2)^2 |\int d\varphi f_\nu^L (f_{\nu_1}^R)^* |^2_{\rho=d/2} N_L^{-1} N_R^{-1} \qquad \text{(B.2)}$$

where d is the distance between the clusters centers, $f^{L(R)}$ are defined by Eq.(7), and $N_{L(R)} = \int d\vec{r}\, |f_{\nu(\nu')}^{L(R)}|^2$ ; the angle $\varphi$ is in the plane perpendicular to the line connecting the centers and located in the middle. As an example, let us write down the expression for the matrix element T(l,m,n;l,0,n) which can be obtained with use of Eqs.(B.1) and (B.2):

$$T_{l,m,n}^{l,0,n} = \frac{8\pi^2 E_0 (ka)^2}{ma^6 (\delta U_0)^2} \frac{|Y_l^m(\theta_0,\varphi_0)|^2 |Y_l^0(\pi)|^2 K_{l+1/2}^4 (pd/2)}{\{(l+1)(pa)^{-1} + (K_{l-1/2}(pa)/K_{l+1/2}(pa)\}^4 K_{l+1/2}^4(pa)} \qquad \text{(B.3)}$$

$\theta_0$ is the angle between the orientations of the clusters axis.

Figure caption:

Fig.1. Nano-based Josephson junction (Color online)

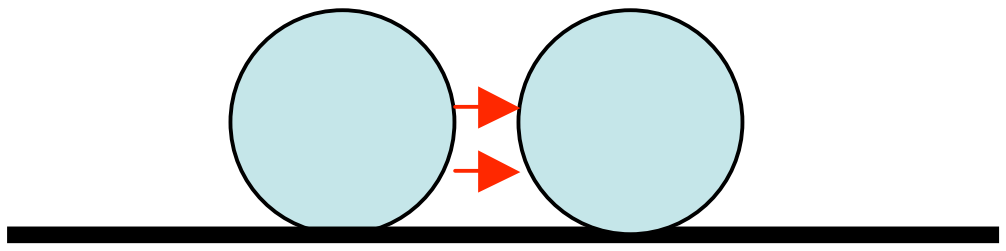